\documentclass[a4paper,11pt,twoside]{article}

\usepackage{geometry} 
\usepackage[latin1]{inputenc}
\usepackage{amsfonts}
\usepackage{amsmath}
\usepackage{amssymb}
\usepackage{amsthm}
\usepackage{graphicx}
\usepackage{booktabs}
\usepackage{accents}
\usepackage{color}
\usepackage{cite}
\usepackage[footnotesize]{caption}

\begin{document}

\begin{titlepage}

\vspace*{-15mm}
\begin{flushright}
MPP-2011-141\\
SISSA 62/2011/EP
\end{flushright}
\vspace*{0.7cm}

\begin{center}
{
\bf\LARGE
Naturalness and GUT Scale Yukawa Coupling Ratios in the CMSSM}
\\[8mm]
Stefan~Antusch$^{\star, \dagger}$
\footnote{E-mail: \texttt{stefan.antusch@unibas.ch}},
Lorenzo~Calibbi$^{\dagger}$
\footnote{E-mail: \texttt{calibbi@mppmu.mpg.de}},
Vinzenz~Maurer$^{\star}$
\footnote{E-mail: \texttt{vinzenz.maurer@unibas.ch}},
Maurizio~Monaco$^{\ddag}$
\footnote{E-mail: \texttt{mmonaco@sissa.it}},
Martin~Spinrath$^{\ddag}$
\footnote{E-mail: \texttt{spinrath@sissa.it}}
\\[1mm]
\end{center}
\vspace*{0.50cm}
\centerline{$^{\star}$ \it
 Department of Physics, University of Basel,}
\centerline{\it
Klingelbergstr.~82, CH-4056 Basel, Switzerland}
\vspace*{0.2cm}
\centerline{$^{\dagger}$ \it
Max-Planck-Institut f\"ur Physik (Werner-Heisenberg-Institut),}
\centerline{\it
F\"ohringer Ring 6, D-80805 M\"unchen, Germany}
\vspace*{0.2cm}
\centerline{$^{\ddag}$ \it
SISSA/ISAS and INFN,}
\centerline{\it
Via Bonomea 265, I-34136 Trieste, Italy }
\vspace*{1.20cm}
\begin{abstract}

\noindent
We analyse the fine-tuning in the Constrained Minimal Supersymmetric Standard Model (CMSSM) in the light of the present and expected ATLAS and CMS SUSY searches. 
Even with 10 fb$^{-1}$ of data and no discovery of SUSY valid regions might remain with fine-tuning less than 20.
Moreover we investigate the fine-tuning price of GUT scale Yukawa coupling relations. Considering a $2\sigma$ constraint for 
$(g-2)_\mu$ and fine-tuning less than 30 yields an allowed
range of $y_\tau/y_b = [1.31,1.70] $, which points towards the alternative GUT prediction
$y_\tau/y_b = 3/2$. Relaxing the $(g-2)_\mu$ constraint to
$5\sigma$ extends the possible region to [1.02,1.70], allowing for approximate $b-\tau$ Yukawa coupling unification. 

\end{abstract}

\end{titlepage}

\section{Introduction}
\addtocounter{footnote}{-5}

The LHC has been colliding protons since more than one and a half year with $\sqrt{s}=7$ TeV and increasing luminosities. So far, about 5 fb$^{-1}$ of integrated luminosity
have been collected and analyses based on $1\div 2$ fb$^{-1}$ of data have been presented by the experimental collaborations in the summer conferences and published
afterwards. Even though no evidences for the Higgs boson nor for new physics (NP) have been found so far, the $\sqrt{s}=7$ TeV run has been very successful and
started to provide relevant constraints on NP models, in particular on low-energy supersymmetry (SUSY). Among those, the most stringent ones are the CMS~\cite{CMS} and ATLAS~\cite{ATLAS} searches for squarks and gluinos in the missing transverse energy plus multi-jets channel, based on $\sim 1$ fb$^{-1}$ of data. These analyses provide bounds on first and second generation squarks ($m_{\tilde q}$) and gluino ($m_{\tilde g}$) masses of about 1 TeV, if 
$m_{\tilde q}\simeq m_{\tilde g}$, and a bound on the gluino mass, $m_{\tilde g}\gtrsim 600$ GeV, if $m_{\tilde q}\gg m_{\tilde g}$~\cite{CMS,ATLAS}.
The prospects are that at least 10~fb$^{-1}$ will be collected before the long shutdown planned in 2013 with the possibility, still under discussion, 
of increasing the center of mass energy to 8 TeV in the 2012 run. 

One main motivation for introducing low-energy SUSY is provided by the stabilization of the electroweak scale. Such a ``natural'' solution of the
hierarchy problem, which requires a SUSY spectrum not far above the electroweak scale, has been already challenged by the negative results of 
SUSY and Higgs boson searches at LEP~\cite{LEP}. However, the degree of fine-tuning required for SUSY to give the correct $Z$ mass depends 
on the actual SUSY model and can be relaxed in specific models that modify the impact of the LEP bound on the Higgs mass (see, for instance,~\cite{NMSSM}). 
Therefore, it is interesting to understand the status of the fine-tuning problem now that the recent SUSY searches at the LHC started to constrain directly the masses 
of the superpartners far beyond the previous LEP and Tevatron bounds. This important question has been recently addressed by a number of papers~\cite{LHC-fine-tuning, arXiv:1110.6926}.
A low degree of fine-tuning is a valuable criterium for discriminating among different SUSY models and, within a given model, to identify
the preferred ranges of the fundamental parameters. 

The only experimental evidence that seems to point towards light NP is the long-standing discrepancy between the Standard
Model (SM) prediction and the experimental determination of the anomalous magnetic moment of the muon, $(g-2)_\mu$, that is currently above the 3$\sigma$ level~\cite{g-2}.
The requirement of a sizeable SUSY contribution to $(g-2)_\mu$ clearly selects a limited portion of the parameter space corresponding, in particular, to quite light
electroweak gauginos and sleptons and sizeable values of $\tan\beta$.\footnote{For a discussion of $(g-2)_\mu$, as well as the interplay of LHC and flavour observables, in
the view of the recent SUSY searches at the LHC, see \cite{arXiv:1111.0176}.} Within models giving definite relations among different superpartner masses, such
as the CMSSM, this translates into the requirement of a generically light SUSY spectrum. The compatibility of the parameter space preferred by $(g-2)_\mu$
with the LHC bounds and a low fine-tuning is an interesting question we are going to address in this paper. 

Besides the interplay of $(g-2)_\mu$ with natural SUSY, the main focus of the present work is on the impact of LHC SUSY searches on 
the GUT scale ratios of the Yukawa couplings and their fine-tuning price. 
The predictions for high-energy Yukawa relations are known to have a sensitive dependence on low-energy SUSY threshold corrections and thus on 
the SUSY spectrum. Such dependence allows us to identify regions of the parameter space preferred by given Yukawa relations and, on the other hand, it might provide
a powerful tool to test high-energy physics, such as
GUT or fermion mass models, once signals of SUSY would be discovered at the LHC. For recent examples of
such interplay, see~\cite{Altmannshofer:2008vr,Antusch:2011sq,maurizio-martin}. In the view of the present and future SUSY searches at the LHC, it can be of interest to check whether peculiar Yukawa GUT ratios, and thus high-energy SUSY scenarios, are favoured by naturalness considerations. 
In the following, we are going to study
this problem within the usual CMSSM, which is a convenient benchmark model providing a spectrum that is approximately realised in many SUSY scenarios, such as 
mSUGRA combined with the assumption of sequestering \cite{Chung:2003fi},
several GUT and fermion mass models with non-Abelian family symmetries, and it is phenomenologically motivated by the so-called SUSY flavour problem, which points towards family universal soft SUSY breaking terms (see, for instance, \cite{Antusch:2011sq} and references therein). The consequences of relaxing the universality assumption, for instance of the Higgs mass parameters, which is known to have an impact on Yukawa unification, see e.g.\ \cite{Altmannshofer:2008vr}, will be studied elsewhere \cite{Antusch:2012xx}.

The rest of the paper is organised as follows: in section 2 we briefly review the used fine-tuning measure and identify low fine-tuning regions of the parameter space. We show the connection of the latter to the GUT $b-\tau$ Yukawa coupling ratios in section 3 before proceeding with a numerical analysis in section 4. Finally, section 5 is devoted to summarize the results and draw conclusions out of them.

\section{Fine-Tuning} \label{sec:FineTuning}

The problem of fine-tuning in the MSSM has been extensively discussed in the literature starting with \cite{Barbieri:1987fn} (for a more extensive list of references, see \cite{arXiv:1110.6926}): here we just present a short review of the problem in order to introduce on wider grounds our analysis and comment on the corrections on the usual tree-level fine-tuning measures that we have included \cite{Cassel:2009ps}.

In the MSSM the Z-boson mass can be expressed in terms of the supersymmetric $\mu$ parameter and the soft masses of the up- and down-type Higgs by means of the tree-level formula 
\begin{equation}
    \frac{M_Z^2}{2} = -|\mu|^2 + \frac{1}{2} \tan 2\beta (m^2_{H_u} \tan\beta - m^2_{H_d} \cot \beta) \;. \label{eq:Zbosonmass}
\end{equation}
The latter, in the case of large $\tan \beta$ in which we are mainly interested, can be expressed as 
\begin{equation}
    \frac{M_Z^2}{2} = -|\mu|^2 - m^2_{H_u} + \mathcal{O}(m^2_{H_{u,d}} / (\tan\beta)^2) \;. \label{eq:Zbosonmass_expansion}
\end{equation}
It is easy to spot that the value of the low energy observable $M_Z$ can thus be obtained by the algebraic sum of the RG evolved GUT parameters $\mu$ and $m^2_{H_u}$, that are quantities of order of the SUSY breaking scale. It is thus evident that a certain amount of tuning is needed to get the correct value of the $Z$ boson mass. 
To quantify the amount of fine-tuning needed in a certain point of the MSSM parameter space, 
the following measure has been introduced \cite{Barbieri:1987fn}
\begin{equation}
    \Delta_a = \left| \frac{\partial \log M_Z}{\partial \log a} \right| = \left|\frac{a}{2 M_Z^2} \frac{\partial M_Z^2}{\partial a} \right| \;.
\end{equation}
$\Delta_a$ reflects the dependence of $M_Z$ on the variation of a given GUT scale Lagrangian parameter $a$.
The value of such a quantity gives exactly what we are interested in: for instance $\Delta_a = 100$ implies a necessary cancellation of $1$ part in $100$. 
The global measure of fine-tuning in a certain point of the parameter space is then simply defined as the maximum of all the single fine-tunings
\begin{equation}
    \Delta = \max_{\substack{a}} \Delta_a \;. \label{eq:wholetuning}
\end{equation}
Now we move to discuss the measure for the single parameters. Setting aside RG effects for the moment, one can easily obtain from eq.~\eqref{eq:Zbosonmass_expansion} the fine-tuning in $\mu$
\begin{equation}
    \Delta_\mu = 2 \frac{\mu^2}{M_Z^2}  \;. \label{eq:mutuning}
\end{equation}
Anyway, $\mu$ is not on the same ground as the other input parameters of the theory, since it is usually determined at low energies so that it yields the correct value for the $Z$ boson mass. One can thus recast $\Delta_\mu$ in terms of $m^2_{H_u}$ as 
\begin{equation}
    \Delta_\mu \approx \left| 2 \frac{m^2_{H_u}}{M_Z^2} + 1 \right| \,.
\end{equation}
From Eq.~\eqref{eq:Zbosonmass_expansion} it is evident that it is necessary to give a closer look at the low energy value of $m^2_{H_u}$, that depends in a non trivial way on the CMSSM parameters $m_0$, $M_{1/2}$ and $A_0$. The latter enter the theory at the GUT scale $M_{\text{GUT}} \approx 2 \cdot 10^{16}$~GeV.

We can express $m^2_{H_u}$ as a polynomial in the CMSSM parameters $m_0$, $M_{1/2}$ and $A_0$ where the coefficients depend on the SUSY scale, the Yukawa and the gauge couplings. Let us now consider as an example the best-fit CMSSM point obtained in~\cite{Buchmueller:2011sw}, with $m_0 = 450$~GeV, $M_{1/2} = 780$~GeV, $A_0 = -1.1$~TeV, $\tan \beta = 41$ and positive $\mu$. For this point we find at the SUSY scale $M_{\text{SUSY}} \approx 1.2$~TeV
\begin{align}
   m^2_{H_u}(M_{\text{SUSY}}) &= 0.058 m_0^2 - 0.094 A_0^2 + 0.317 A_0 M_{1/2} -1.304 M_{1/2}^2 \;.
\end{align}
The form of the polynomial can be understood considering one-loop RG equation of $m^2_{H_u}$. The coefficients were computed by fitting such a polynomial in a neighborhood of the best-fit point and cross-checked with the semi-analytic formula obtained similarly to \cite{Codoban:1999fp}.
The coefficients in this equation depend strongly on the top Yukawa coupling and the strong coupling
constant. Therefore it is important to consider the fine-tuning also in this quantities \cite{hep-ph/9303291}, weighting them properly by their experimental uncertainty. Anyway they turn out to be subleading and hence we do not discuss them further.

Due to the presence of the term $A_0 M_{1/2}$ in the polynomial the two parameters cannot be treated independently in the determination of the fine-tuning. In turn we consider
\begin{equation}
\begin{split}
    \Delta_{A_0, M_{1/2}} &= \max \{ \Delta_{A_0}, \Delta_{M_{1/2}} \} \\ 
    &= \left| \frac{M_{1/2}^2}{M_Z^2} \right| \max \{ |- 0.19 a_0^2 + 0.32 a_0|, |- 2.61 + 0.32 a_0| \} =: \left| \frac{M_{1/2}^2}{M_Z^2} \right| \rho(a_0) \;, \label{eq:AMtuning}
\end{split}
\end{equation}
where $a_0 \equiv A_0 / M_{1/2}$. Notice that large gaugino masses as well as large trilinear couplings give large fine-tuning. Another feature is noteworthy, showing the close connection of $A_0$ and $M_{1/2}$. Given a specific value for $M_{1/2}$ the minimal fine-tuning is obtained for $\hat{a}_0 \approx 3.7$ and not as naively expected for $a_0 = 0$, yielding $\rho(\hat{a}_0) \approx 1.43$ compared to $\rho(0) \approx 2.61$. This is due to the large negative contribution 
in the RGE running of the stop trilinear coupling $A_t$ provided by gluino-top loops, that usually drives $A_t$ to large and negative low-energy values even with
a vanishing inital value at the GUT scale.

The discussion so far was very instructive since it gives a general flavour of the main features of the fine-tuning measure, but it has to be refined to take into account loop effects \cite{Cassel:2009ps}, which modify the original eq.~\eqref{eq:Zbosonmass}. The effect of going beyond tree-level is two-fold. On one side there is an obvious modification to the Higgs potential; on the other side loop corrections affect also the Higgs vacuum expectation value $v$. We do not go into detailed calculations here, but we just notice that taking into account all the relevant corrections roughly reduces the fine-tuning by a factor $M_Z^2/m_h^2$, where $m_h$ is the light Higgs boson mass. Such effect is clearly sizeable in the parameter region of interest, where $m_h \gtrsim 110$~GeV. In turn considering the one-loop effects shifts the lowest fine-tuning regions given in terms of $\hat{a}_0$, because of the Higgs mass dependence on the trilinear soft terms\footnote{In fact our full numerical analysis finds a minimum of the fine-tuning for $\hat{a}_0 \approx 3$}.

Fine-tuning measures in general should be taken with a grain of salt. It is a matter of personal taste how much fine-tuning one accepts as ``natural'' and, clearly, 
changing the fine-tuning measure changes as well the ``amount'' of fine-tuning. Nevertheless, we want to stress here that even in the CMSSM there are still regions of the allowed parameter space with a reasonable amount of fine-tuning, rendering them more attractive than others.

\section{Connection to the GUT $\boldsymbol{b-\tau}$ Yukawa Coupling Ratio}
\label{sec:yukawas}

It is well known that in the MSSM exists a class of one-loop
corrections to the Yukawa couplings of the down-type quarks and the charged
leptons, which are sizeable if $\tan \beta$ is large \cite{SUSYthresholds}. 
These corrections obviously depend on the SUSY breaking parameters, 
which can be constrained by the requirement of low
fine-tuning, as discussed in the last section. Here we estimate
the size of the SUSY threshold corrections in the low fine-tuning region.

We focus here on the third generation: for the bottom quark mass we can write
\begin{equation}
m_b = y_b v_d (1 + \epsilon_b \tan \beta) \;,
\end{equation}
where, neglecting bino and wino contributions and EWSB effects, the correction $\epsilon_b$ is given by \cite{Freitas:2007dp, Antusch:2008tf}
\begin{equation} \label{eq:epsilonb}
\epsilon_b \approx \epsilon_0 + \epsilon_Y y_t^2 = - \frac{2}{3 \pi} \alpha_s \frac{\mu}{M_3} H_2\left(\tfrac{m_{\tilde{Q}_3}^2}{M_3^2}, \tfrac{m_{\tilde{d}_3}^2}{M_3^2}\right) - \frac{y_t^2}{16 \pi^2}  \frac{A_t}{\mu} H_2\left(\tfrac{m_{\tilde{Q}_3}^2}{\mu^2}, \tfrac{m_{\tilde{u}_3}^2}{\mu^2}\right) \;,
\end{equation}
where the loop function $H_2$ is defined as
\begin{equation}
    H_2(x, y) = \frac{x \ln x}{(1-x)(x-y)} + \frac{y \ln y}{(1-y)(y-x)} \;.
\end{equation}
The correction $\epsilon_b$ becomes bigger if $M_3$ and $A_t$
increase and the sfermion mass parameters decrease. The restrictions derived
from low fine-tuning are strong only for the gluino mass
and the soft trilinears: from the tree-level considerations of the previous
section indeed we know that small $M_{1/2}$ and
$a_0 \approx 3.7$ are preferred. The dependence of the fine-tuning on $m_0$ and
$\tan \beta$ (for $\tan \beta \gtrsim 10$) is rather weak, while that on $\mu$ is stronger.
Anyway, $\mu$ itself is not
a free parameter in our setup, being determined from successful EWSB.

To estimate the size of the SUSY threshold corrections we set therefore
$a_0 = 3.7$, $\tan \beta = 40$, $m_0 = 1.5$~TeV,  $M_{1/2} = 350$~GeV.
The choice for $\tan \beta$ was inspired again by the best-fit point of~\cite{Buchmueller:2011sw}, while $m_0$ and $M_{1/2}$ were chosen such that 
the recent collider bounds \cite{CMS,ATLAS} are satisfied and the resulting fine-tuning is not very large. From our numerical analysis including loop corrections to the fine-tuning measure we find for this point a fine-tuning
of about 25 for either sign of $\mu$.

Using {\tt softSUSY} \cite{Allanach:2001kg} for the calculation of the input parameters of Eq.\ \eqref{eq:epsilonb}, we find
\begin{align}
\epsilon_b = \begin{cases} +2.2 \times 10^{-3} & \text{for } \mu > 0 \;, \\  -2.3 \times 10^{-3} & \text{for } \mu < 0 \;. \\ \end{cases}
\end{align}
At the GUT scale $y_\tau/y_b \approx 1.26 (1+ \epsilon_b \tan \beta)$
\cite{Antusch:2008tf, Antusch:2011sq} and hence we find, for the considered point, 
$y_\tau/y_b \approx 1.4$ for $\mu > 0$ and $y_\tau/y_b \approx 1.1$
for $\mu < 0$. This already tells us that SUSY threshold
corrections to the Yukawa coupling ratios for the low fine-tuning region are in the ballpark to correctly reproduce both the common GUT prediction $y_\tau/y_b = 1$
as well as the new relation $y_\tau/y_b = 3/2$ \cite{arXiv:0902.4644}.

To perfectly reproduce
these predictions we would prefer slightly larger threshold
corrections, which can be accomodated by increasing
$M_{1/2}$, even though this would make the fine-tuning larger. 
These considerations are nevertheless based on the tree-level
relations of the previous section; including loop corrections
to the fine-tuning could have some impact on the
``natural'' values of the SUSY threshold corrections. Furthermore considering 
experimental results will constrain them even more 
and will rule out, for instance, the $\mu < 0$ case as we will see in the following section.

\section{Numerical Analysis}
\label{sec:results}

\begin{figure}
\centering
\includegraphics[width=0.85\textwidth]{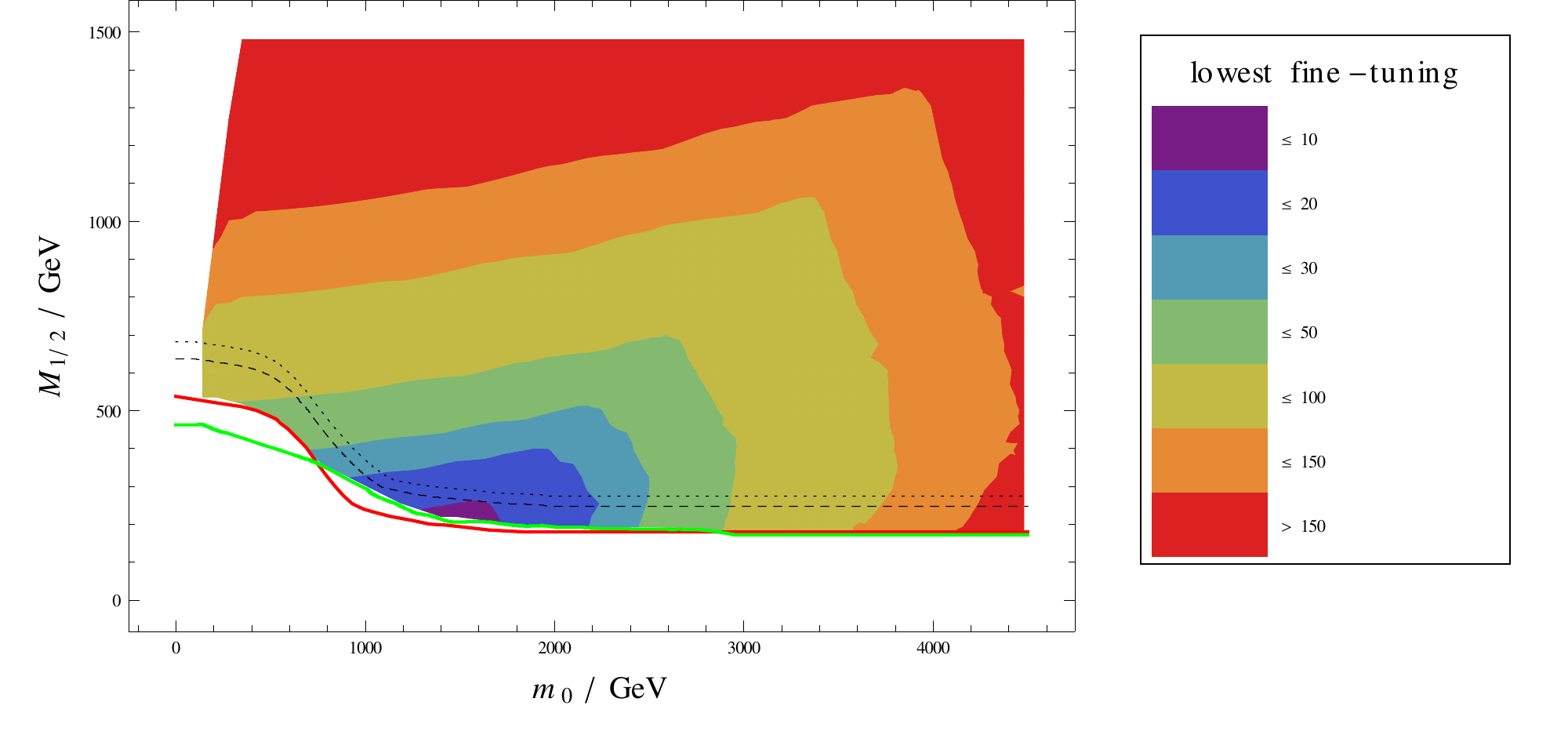}
\includegraphics[width=0.85\textwidth]{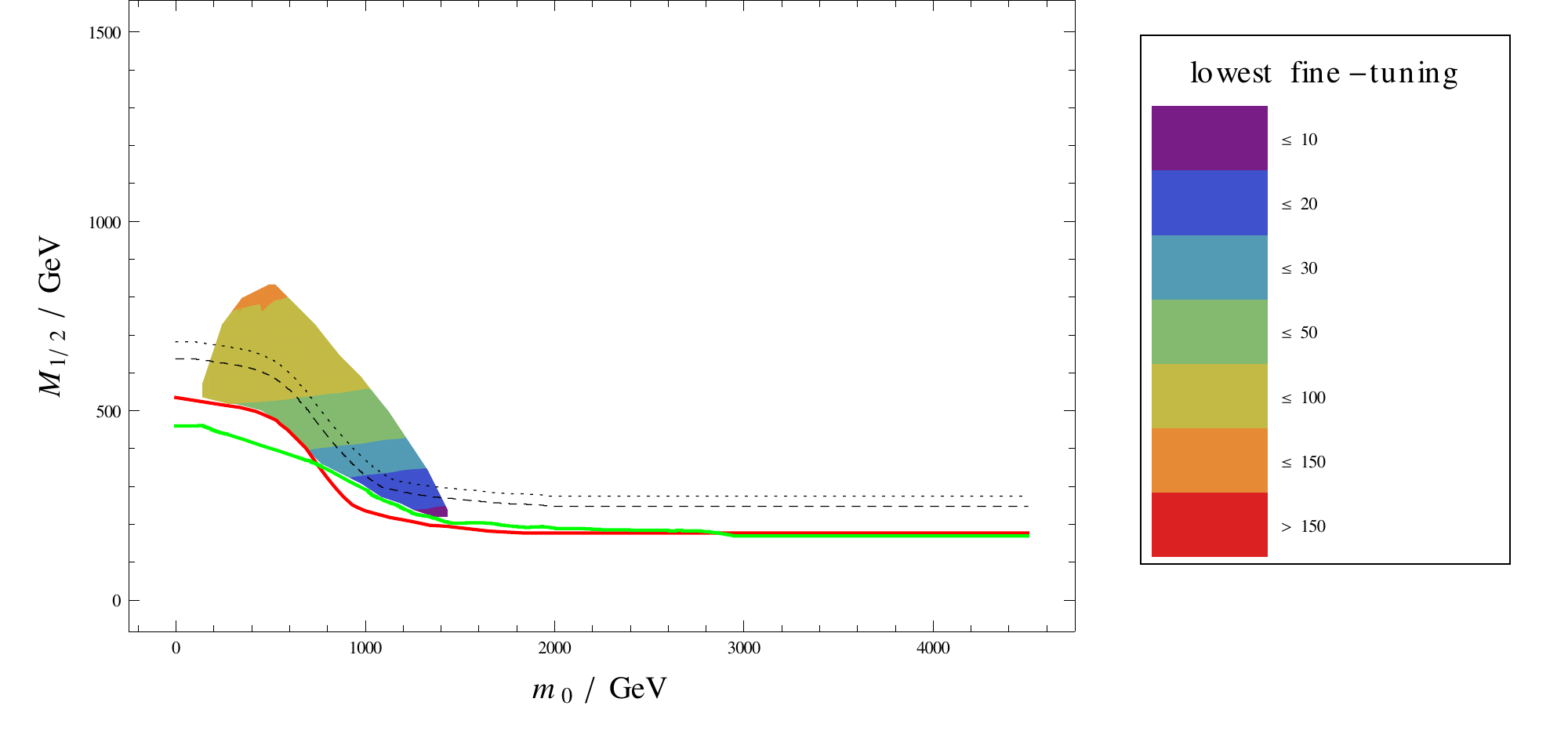}
\caption{
Lowest fine-tuning in the $m_0$-$M_{1/2}$ plane consistent 
with current bounds on $b \to s \gamma$, $B_s \to \mu^+ \mu^-$,
collider bounds and $(g-2)_\mu$ at 5 (2)$\sigma$ in the upper (lower)
plot. 
$a_0 = A_0/M_{1/2}$ has been varied from $-20$ to 20 and $\tan \beta$ 
from 10 to 60 (for details, see main text). 
The green and red lines correspond to current CMS \cite{CMS}
and ATLAS \cite{ATLAS} bounds. The black dashed (dotted) line
represents the extrapolated CMS exclusion line for 5~fb$^{-1}$
(10~fb$^{-1}$) of collected data.
}\label{fig:M12overm0}
\end{figure}

\begin{figure}
\centering
\includegraphics[width=0.85\textwidth]{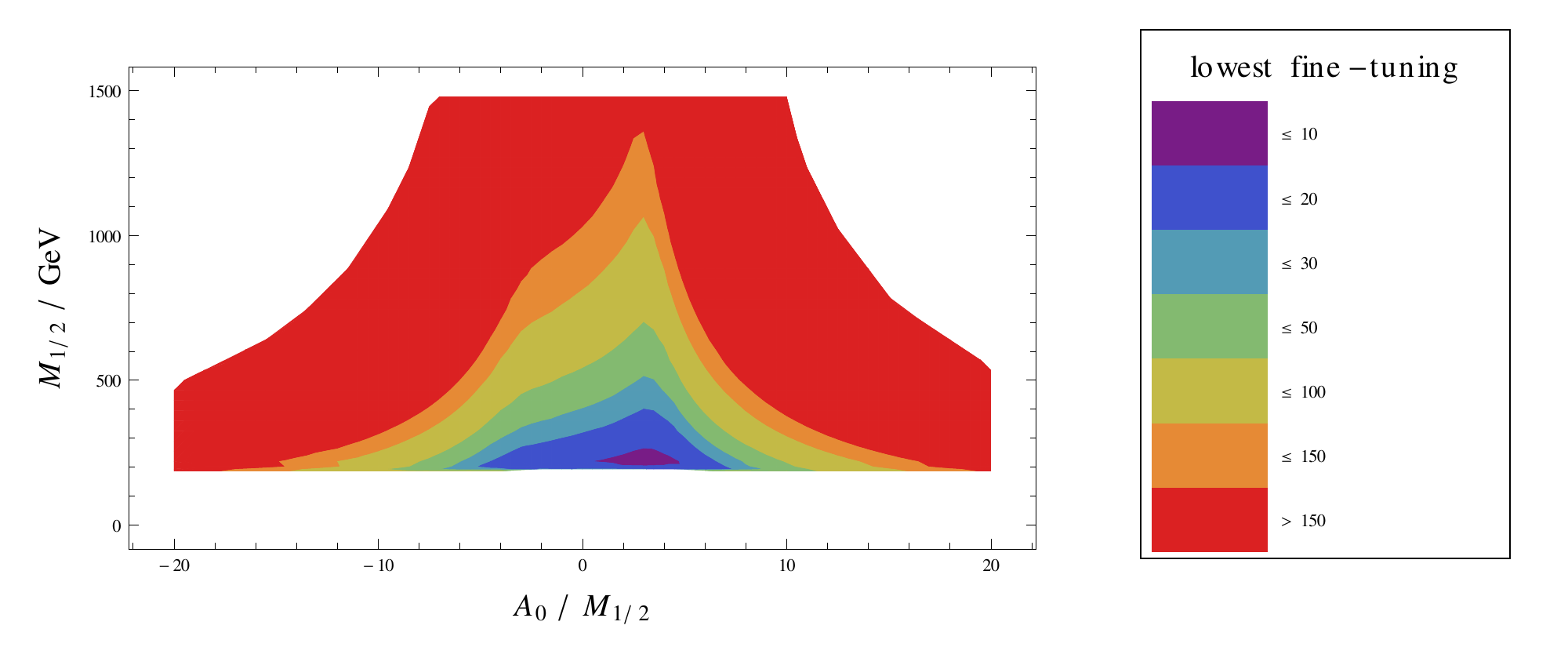}
\includegraphics[width=0.85\textwidth]{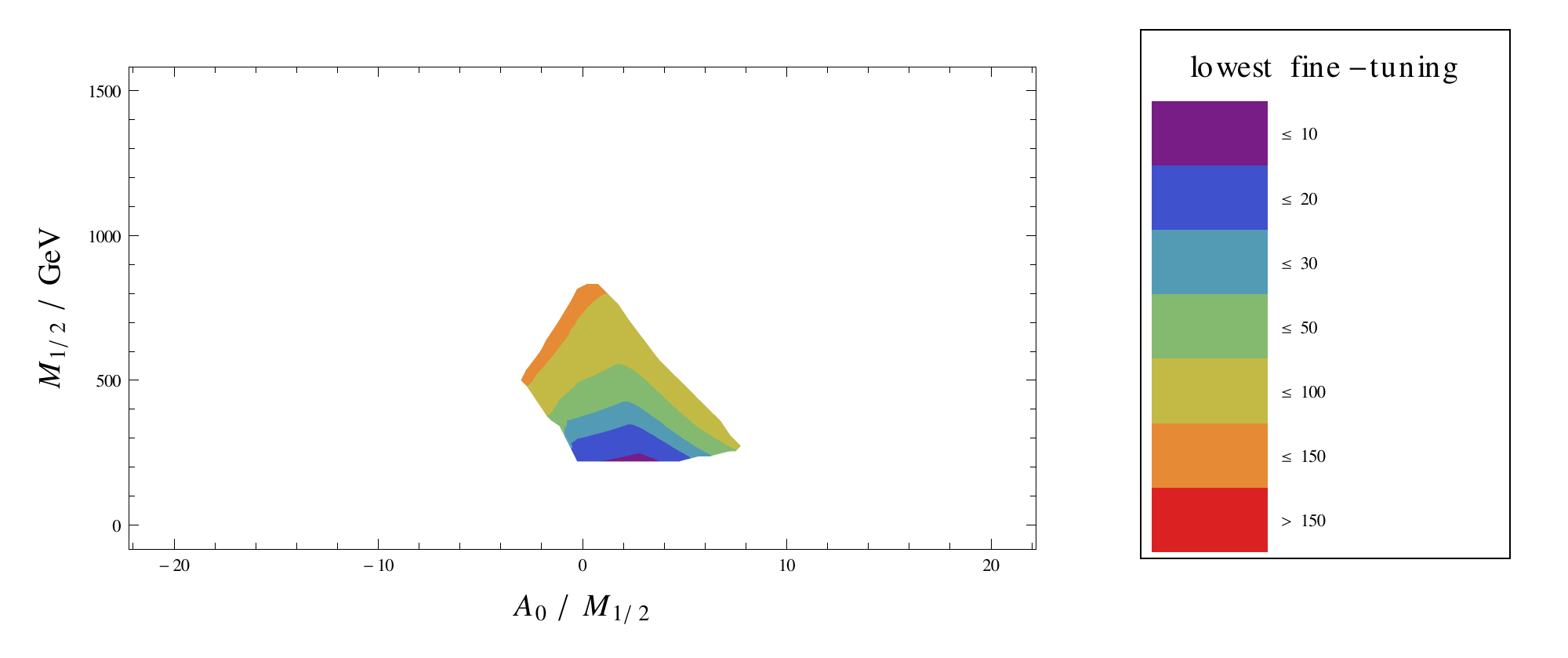}
\caption{
Lowest fine-tuning in the $A_0/M_{1/2}$-$M_{1/2}$ plane consistent 
with current bounds on $b \to s \gamma$, $B_s \to \mu^+ \mu^-$,
collider bounds and $(g-2)_\mu$ at 5 (2)$\sigma$ in the upper (lower)
plot. 
$m_0$ has been varied from $0$ to $4.5$ TeV and $\tan \beta$ from $10$ to $60$ (for details, see main text). 
}\label{fig:M12overa0}
\end{figure}

\begin{figure}
\centering
\includegraphics[width=0.85\textwidth]{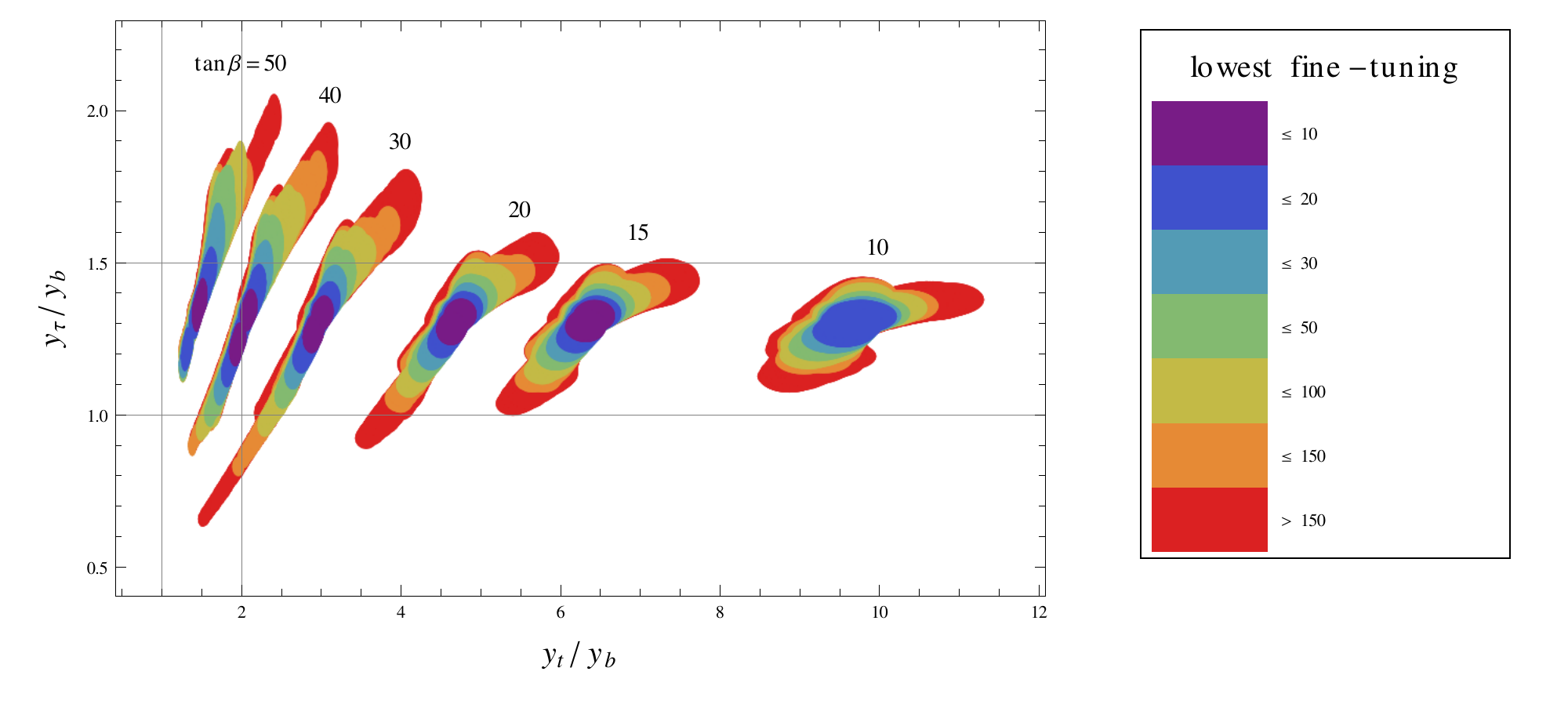}
\includegraphics[width=0.85\textwidth]{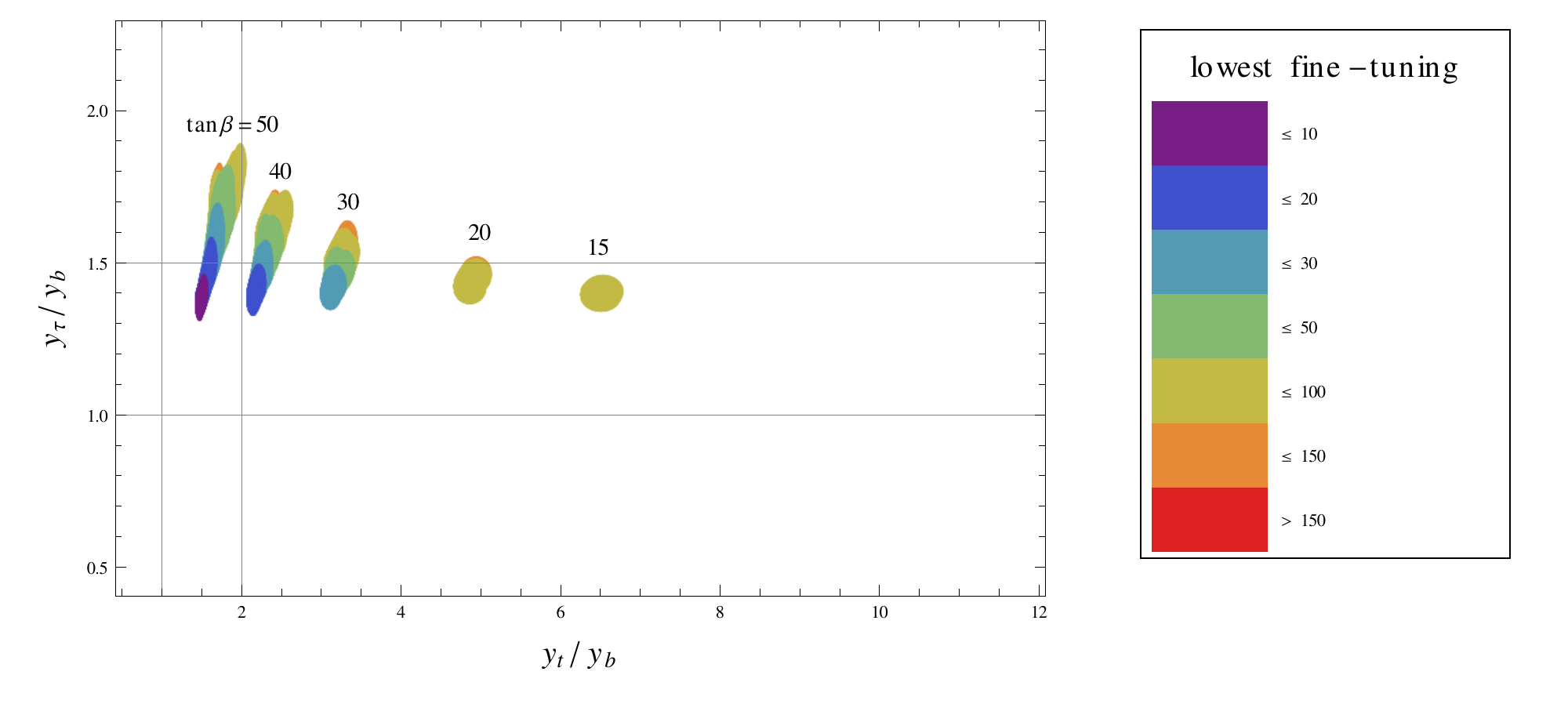}
\caption{
Lowest fine-tuning in the GUT scale $y_\tau / y_b$-$y_t / y_b$ plane
consistent with current bounds on $b \to s \gamma$, $B_s \to \mu^+ \mu^-$,
collider bounds and $(g-2)_\mu$ at 5 (2)$\sigma$ in the upper (lower)
plot for $\tan \beta = 10,$ 15, 20, 30, 40, 50 from right to left. For $\tan \beta = 60$ there is no
region allowed in our scan.
The 1$\sigma$ errors on the quark masses \cite{PDG} are taken into account
by scaling the data points correspondingly.
}\label{fig:Yukawas}
\end{figure}

To calculate the SUSY spectra, GUT scale Yukawa couplings and
the fine-tuning, we used a modified version of
{\tt softSUSY} \cite{Allanach:2001kg}. The modifications take
into account the implicit $M_Z$ dependence on the Higgs vev \cite{Cassel:2009ps} and the SUSY threshold
corrections for all three fermion generations.

For the calculation of the observables $(g-2)_\mu$, BR$(b \to s \gamma)$,
BR$(B_s \to \mu^+ \mu^-)$ and the Higgs boson mass 
we used {\tt SuperIso} \cite{Mahmoudi:2008tp}.
The experimental values for the observables above are: $\delta a_\mu = a_\mu^{\text{exp}} - a_\mu^{\text{SM}} = (2.61 \pm 0.8) \cdot 10^{-9}$ \cite{g-2}, where $a_\mu = (g-2)_\mu/2$, BR$(b \to s \gamma) = (355 \pm 24 \pm 9)\cdot 10^{-6}$ \cite{arXiv:1010.1589},
BR$(B_s \to \mu^+ \mu^-) < 1.08 \cdot 10^{-8}$ at $95\%$ CL \cite{CMS-PAS-BPH-11-019}. In our numerical analysis we allow for $95\%$ CL regions in which we consider also theoretical uncertainties. 
Additionally, LEP bounds \cite{PDG} and current CMS
and ATLAS SUSY exclusion bounds based on roughly 1 fb$^{-1}$ of data
\cite{CMS,ATLAS} were applied. Furthermore we discard points corresponding to
a charged LSP.\footnote{We do not impose any bound on neutralino relic density,
nor from direct and indirect dark matter searches. For recent discussions of these constraints, 
see \cite{Profumo:2011zj,Buchmueller:2011ki} and, in the context of natural SUSY scenarios, the last two papers in Ref. \cite{LHC-fine-tuning}.}

Due to the ongoing debate on the status of $(g-2)_\mu$ 
we discuss results for two different bounds. Firstly we are
conservative and demand that the prediction for $(g-2)_\mu$
should not deviate more than 5$\sigma$ from the experimental
result \cite{g-2}. In this case there is still room for SUSY corrections
to increase the tension between the SM and experiments, in particular allowing
points with a negative $\mu$ parameter. In a second
step we require SUSY to alleviate the tension and show only regions
where it is less than 2$\sigma$. This does not only exclude
$\mu < 0$, but also gives effectively an upper bound on the sparticle
masses.

We scanned the CMSSM parameter space extensively, $m_0$
from 0 to 4.5~TeV, $M_{1/2}$ from 0.15 to 1.5~TeV, and
$a_0 = A_0/M_{1/2}$ from $-20$ to 20  for $\tan \beta = 10$, 15, 20,
30, 40, 50, 60.
We also scanned over both signs of $\mu$.
Our results are presented in Figs.\ \ref{fig:M12overm0}-\ref{fig:Yukawas} and Tab.\ \ref{tab:Ranges}.
In Figs.\ \ref{fig:M12overm0} and \ref{fig:M12overa0} we show respectively the
fine-tuning in the $m_0$-$M_{1/2}$ and $a_0$-$M_{1/2}$ planes for
5 and 2$\sigma$ allowed tension in $(g-2)_\mu$. In both cases
we have shown for each point in the planes the lowest fine-tuning obtained by variation
of the remaining parameters. In Fig.\ \ref{fig:M12overm0}, we show the present CMS and ATLAS
exclusion curves as red and green lines respectively \cite{CMS, ATLAS}. Since they are based on SUSY searches that
are sensitive to gluino and first/second generation squarks only, whose masses are almost insensitive to
variations of $A_0$ and $\tan\beta$, we can safely consider these LHC bounds to constrain the $m_0$-$M_{1/2}$ 
plane only and varying the other parameters. The black dashed and dotted lines show our estimate for
the LHC exclusion potential with 5 and 10 fb$^{-1}$ respectively. They were obtained from
an extrapolation of the CMS curve based on an analysis of the total production cross-section of coloured SUSY particles 
at LHC with $\sqrt{s}=7$ TeV by means of {\tt PROSPINO 2.0}~\cite{prospino}. Of course, they represent
just a naive approximation meant to be indicative of the 2012 LHC sensitivity.

First of all it is interesting to note that after the first one and a half
year of LHC there are still regions in the CMSSM allowed with 
fine-tuning less than 10. It is also interesting, that
even after 10 fb$^{-1}$ with negative results in the SUSY searches,
and imposing the restrictive 2$\sigma$ bound on $(g-2)_\mu$,
a parameter region with fine-tuning less than 20 would still be
allowed.

In Fig.\ \ref{fig:M12overa0} we also see that a value of
$a_0 \approx 3$ reduces the fine-tuning significantly.
This value is a bit smaller than the one we derived in section
\ref{sec:FineTuning}, but still confirms our findings.
Values much larger or smaller than this easily have a
fine-tuning larger than 100. From this point of view the
solution for $b$-$\tau$ Yukawa coupling unification in
the CMSSM with small gaugino masses, positive $\mu$ and large values
for $m_0$ and $|A_0|$ (see, e.g.~\cite{Altmannshofer:2008vr}) is clearly disfavoured.

\begin{table}
\centering
\begin{tabular}{c cc c cc}
\toprule
           & \multicolumn{2}{c}{$\delta_\mu \leq 5\sigma$} & \phantom{a} & \multicolumn{2}{c}{$\delta_\mu \leq 2\sigma$} \\ \cmidrule{2-3}  \cmidrule{5-6}
  $\Delta$ & $y_\tau/y_b$ & $y_\mu/y_s$ && $y_\tau/y_b$ & $y_\mu/y_s$  \\ \midrule
   $< 10$ & [1.16,  1.47] & [2.98,  5.53] && [1.31,  1.47] & [3.25,  5.53]\\
   $< 20$ & [1.07,  1.59] & [2.83,  5.89] && [1.31,  1.59] & [3.25,  5.89]\\
   $< 30$ & [1.02,  1.70] & [2.70,  6.21] && [1.31,  1.70] & [3.25,  6.21]\\
   $< 40$ & [0.97,  1.78] & [2.61,  6.38] && [1.31,  1.78] & [3.25,  6.39]\\
   $< 50$ & [0.93,  1.82] & [2.52,  6.49] && [1.31,  1.82] & [3.25,  6.49]\\
\bottomrule
\end{tabular}
\caption{Range of allowed GUT scale Yukawa coupling ratios $y_\tau/y_b$ and $y_\mu/y_s$ including the 1$\sigma$ errors for the quark masses \cite{PDG} for different upper bounds on fine-tuning including collider bounds, the constraints from $B$ physics and the difference $\delta_\mu$ between the theoretical prediction and the measurement of the anomalous magnetic moment of the muon at 5$\sigma$ and 2$\sigma$. \label{tab:Ranges}}
\end{table}

Finally, we comment on the preferred GUT scale Yukawa
coupling relations. In Fig.~\ref{fig:Yukawas} we show the
third generation GUT scale ratios $y_\tau/y_b$ and
$y_t/y_b$ for different values of $\tan \beta$. If we allow a 5$\sigma$ tension in $(g-2)_\mu$, 
we see that both definite GUT predictions $y_\tau/y_b = 1$
and $y_\tau/y_b = 3/2$ can be achieved with a moderate
amount of fine-tuning (with a slight preference for $y_\tau/y_b = 3/2$), 
but the $y_\tau/y_b = 1$ is ruled out
by the 2$\sigma$ bound on $(g-2)_\mu$. This is due to the fact that
$(g-2)_\mu$ strongly disfavour negative values of the $\mu$ parameter. It is also
interesting to note that we do not find unification of the bottom with the
top Yukawa coupling. In fact we scanned $\tan \beta$
up to 60, but this region is completely
ruled out in our scan by the bound on $B_s \to \mu^+ \mu^-$, so that we find
always $y_t/y_b > 1.15$ even if we allow a 5$\sigma$ deviation in $(g-2)_\mu$.
Nevertheless, many studies about $b$-$\tau$ or $t$-$b$-$\tau$ Yukawa
coupling unification, for some examples see \cite{Altmannshofer:2008vr},
allow for non-universalities of the GUT scale Higgs mass parameters, which
might alter the results.

In Tab.~\ref{tab:Ranges} we have collected allowed ranges
of GUT scale Yukawa coupling ratios for different
upper bounds on $\Delta$. Note that for $\Delta < 10$ we find
no agreement neither with $y_\tau/y_b = 1$ nor with
$y_\tau/y_b = 3/2$, although bigger values are preferred
and a deviation from the definite prediction could be
explained, for instance, by mixing effects
\cite{arXiv:1111.3639}. Still the $y_\tau/y_b = 3/2$
ratio can be already obtained with $\Delta < 20$, while
$y_\tau/y_b = 1$ corresponds at least to $\Delta \simeq 40$
and it is not consistent with the $(g-2)_\mu$ tension
below the 2$\sigma$ level.

The situation for the second generation is also noteworthy.
In Tab.~\ref{tab:Ranges} we see that 
values bigger than the famous Georgi--Jarlskog relation
$y_\mu/y_s = 3$ \cite{HUTP-79-A026} are somewhat preferred, see also
 \cite{Antusch:2008tf, arXiv:0902.4644}. Nevertheless,
here experimental errors are quite big and mixing
effects can be expected to be more sizeable than those for the
third generation.

\section{Summary and Conclusions}
\label{sec:conclusion}

In this paper, we have analysed fine-tuning in the Constrained Minimal Supersymmetric Standard Model (CMSSM) and the connection to the allowed GUT scale ratios of the third family Yukawa couplings. We chose the CMSSM as a benchmark since it is approximately realised in many SUSY scenarios, such as several GUT and fermion mass models with non-Abelian family symmetries, and it is phenomenologically motivated by the SUSY flavour problem. 

In order to provide a natural solution to the hierarchy problem, a SUSY model should not require too much tuning of parameters, which is often expressed in terms of a reasonable fine-tuning measure $\Delta$. We have presented a detailed analysis of fine-tuning in an extensive range of CMSSM parameters, and discussed the impact of the present phenomenological constraints as well as of the expected future sensitivities of the ATLAS and CMS SUSY searches. Considering the present results, including $(g-2)_\mu$ at $2\sigma$, we find that the minimal fine-tuning measure $\Delta$ is $7.7$. Even if 10 fb$^{-1}$ of data would not give any hint on SUSY, there might still remain an untested part of the parameter space where $\Delta$ is less than $20$. 

In a bottom-up analysis, we have then investigated the possible GUT scale ratios of the third family Yukawa couplings and their fine-tuning price.
Requiring consistency with present constraints, an alleviated tension between the measured $(g-2)_\mu$ and the theoretical prediction
at the level of $2\sigma$ as well as fine-tuning less than 30, we find an allowed
range of $y_\tau/y_b = [1.31, 1.70]$, which points towards alternatives to $b-\tau$ Yukawa unification such as $y_\tau/y_b = 3/2$ \cite{arXiv:0902.4644}. 
On the other hand, if the $(g-2)_\mu$ constraint is relaxed to 5$\sigma$, the allowed region includes points with a negative $\mu$ parameter and the range for $y_\tau/y_b$ extends to [1.02, 1.70]. Hence,
$b-\tau$ Yukawa unification may be realised with a relatively low fine-tuning price, if one is willing to pay for this in form of a larger deviation in the anomalous magnetic moment of the muon.

In summary, SUSY GUT models with SUSY breaking close to the CMSSM and the prediction $y_\tau/y_b = 3/2$ would still provide a comparatively ``natural" scenario.

\section*{Acknowledgements}

We would like to thank Dean Horton, Sabine Kraml and Graham Ross for useful discussions. S.A.\ and V.M.\ were supported by the Swiss National Science Foundation. S.A.\ acknowledges partial support by the DFG cluster of excellence ``Origin and Structure of the Universe''.


\begin{thebibliography}{99}




\bibitem{CMS}
  S.~Chatrchyan {\it et al.} [CMS Collaboration],
  arXiv:1109.2352 [hep-ex].

\bibitem{ATLAS}
  G.~Aad {\it et al.} [ATLAS Collaboration],
  arXiv:1109.6572 [hep-ex].

\bibitem{LEP}
  L.~Giusti, A.~Romanino and A.~Strumia,
  Nucl.\ Phys.\ B\ {\bf 550} (1999) 3
  [hep-ph/9811386];
  R.~Barbieri and A.~Strumia,
  hep-ph/0007265.

\bibitem{NMSSM}
  R.~Dermisek and J.~F.~Gunion,
  Phys.\ Rev.\ Lett.\ \ {\bf 95} (2005) 041801
  [hep-ph/0502105].


\bibitem{LHC-fine-tuning}
  A.~Strumia,
  JHEP\ {\bf 1104} (2011) 073
  [arXiv:1101.2195 [hep-ph]];
  U.~Ellwanger, G.~Espitalier-Noel and C.~Hugonie,
  arXiv:1107.2472 [hep-ph];
  M.~Asano, T.~Moroi, R.~Sato and T.~T.~Yanagida,
  arXiv:1111.3506 [hep-ph];
  S.~Akula, M.~Liu, P.~Nath and G.~Peim,
  arXiv:1111.4589 [hep-ph];
   S.~Amsel, K.~Freese and P.~Sandick,
  arXiv:1108.0448 [hep-ph].

\bibitem{arXiv:1110.6926}
  M.~Papucci, J.~T.~Ruderman and A.~Weiler,
  arXiv:1110.6926 [hep-ph].

\bibitem{g-2}
  M.~Passera, W.~J.~Marciano, A.~Sirlin,
  Phys.\ Rev.\  {\bf D78 } (2008)  013009.
  [arXiv:0804.1142 [hep-ph]];
  K.~Hagiwara, R.~Liao, A.~D.~Martin, D.~Nomura, T.~Teubner,
  J.\ Phys.\ G {\bf G38}, 085003 (2011).
  [arXiv:1105.3149 [hep-ph]].


\bibitem{arXiv:1111.0176}
  L.~Calibbi, R.~N.~Hodgkinson, J.~Jones-Perez, A.~Masiero and O.~Vives,
  arXiv:1111.0176 [hep-ph].

\bibitem{Altmannshofer:2008vr}
W.~Altmannshofer, D.~Guadagnoli, S.~Raby, D.~M.~Straub,
Phys.\ Lett.\  {\bf B668 } (2008)  385-391.
[arXiv:0801.4363 [hep-ph]];
H.~Baer, S.~Kraml, A.~Lessa, S.~Sekmen,
JHEP {\bf 1002 } (2010)  055.
[arXiv:0911.4739 [hep-ph]];
 I.~Gogoladze, R.~Khalid, Q.~Shafi,
 Phys.\ Rev.\  {\bf D79 } (2009)  115004.
 [arXiv:0903.5204 [hep-ph]];
 I.~Gogoladze, S.~Raza, Q.~Shafi,
 [arXiv:1104.3566 [hep-ph]];
 S.~Dar, I.~Gogoladze, Q.~Shafi, C.~S.~Un,
 [arXiv:1105.5122 [hep-ph]];
  M.~Badziak, M.~Olechowski and S.~Pokorski,
  JHEP\ {\bf 1108}, 147  (2011)
  [arXiv:1107.2764 [hep-ph]];
  M.~A.~Ajaib, T.~Li and Q.~Shafi,
  arXiv:1111.4467 [hep-ph].
  
  
\bibitem{Antusch:2011sq}
  S.~Antusch, L.~Calibbi, V.~Maurer, M.~Spinrath,
  Nucl.\ Phys.\  {\bf B852 } (2011)  108-148.
  [arXiv:1104.3040 [hep-ph]].

\bibitem{maurizio-martin}
  M.~Monaco and M.~Spinrath,
  Phys.\ Rev.\ D\ {\bf 84} (2011) 055009
  [arXiv:1106.6208 [hep-ph]].

\bibitem{Chung:2003fi}
  D.~J.~H.~Chung, L.~L.~Everett, G.~L.~Kane, S.~F.~King, J.~D.~Lykken and L.~-T.~Wang,
  Phys.\ Rept.\  {\bf 407} (2005) 1
  [hep-ph/0312378].

\bibitem{Antusch:2012xx}
  S.~Antusch, L.~Calibbi, V.~Maurer, M.~Monaco, M.~Spinrath,
  work in progress.

\bibitem{Barbieri:1987fn}
  R.~Barbieri, G.~F.~Giudice,
  Nucl.\ Phys.\  {\bf B306 } (1988)  63.

\bibitem{Cassel:2009ps}
  S.~Cassel, D.~M.~Ghilencea, G.~G.~Ross,
  Nucl.\ Phys.\  {\bf B825 } (2010)  203-221.
  [arXiv:0903.1115 [hep-ph]];
  D.~Horton, G.~G.~Ross,
  Nucl.\ Phys.\  {\bf B830 } (2010)  221-247.
  [arXiv:0908.0857 [hep-ph]];
  S.~Cassel, D.~M.~Ghilencea, G.~G.~Ross,
  Phys.\ Lett.\  {\bf B687 } (2010)  214-218.
  [arXiv:0911.1134 [hep-ph]];
  S.~Cassel, D.~M.~Ghilencea and G.~G.~Ross,
  Nucl.\ Phys.\  B {\bf 835} (2010) 110
  [arXiv:1001.3884 [hep-ph]];
  S.~Cassel, D.~M.~Ghilencea, S.~Kraml, A.~Lessa, G.~G.~Ross,
  JHEP {\bf 1105 } (2011)  120.
  [arXiv:1101.4664 [hep-ph]].
  
\bibitem{Buchmueller:2011sw}
  O.~Buchmueller, R.~Cavanaugh, A.~De Roeck, M.~J.~Dolan, J.~R.~Ellis, H.~Flacher, S.~Heinemeyer, G.~Isidori {\it et al.},
  [arXiv:1110.3568 [hep-ph]].
  
\bibitem{Codoban:1999fp}
  S.~Codoban, D.~I.~Kazakov,
  Eur.\ Phys.\ J.\  {\bf C13 } (2000)  671-679.
  [hep-ph/9906256].

\bibitem{hep-ph/9303291}
  B.~de Carlos and J.~A.~Casas,
  Phys.\ Lett.\ B\ {\bf 309} (1993) 320
  [hep-ph/9303291];
  L.~Giusti, A.~Romanino and A.~Strumia,
  Nucl.\ Phys.\ B\ {\bf 550} (1999) 3
  [hep-ph/9811386];
  A.~Romanino and A.~Strumia,
  Phys.\ Lett.\ B\ {\bf 487} (2000) 165
  [hep-ph/9912301].
 
  \bibitem{SUSYthresholds}
 L.~J.~Hall, R.~Rattazzi and U.~Sarid,
 Phys.\ Rev.\  D {\bf 50} (1994) 7048
 [arXiv:hep-ph/9306309];
 M.~S.~Carena, M.~Olechowski, S.~Pokorski and C.~E.~M.~Wagner,
 Nucl.\ Phys.\  B {\bf 426} (1994) 269
 [arXiv:hep-ph/9402253];
 R.~Hempfling,
 Phys.\ Rev.\  D {\bf 49} (1994) 6168;
 T.~Blazek, S.~Raby and S.~Pokorski,
 Phys.\ Rev.\  D {\bf 52} (1995) 4151
 [arXiv:hep-ph/9504364].

\bibitem{Freitas:2007dp}
  A.~Freitas, E.~Gasser and U.~Haisch,
  Phys.\ Rev.\  D {\bf 76} (2007) 014016
  [arXiv:hep-ph/0702267].
  
\bibitem{Antusch:2008tf}
  S.~Antusch, M.~Spinrath,
  Phys.\ Rev.\  {\bf D78 } (2008)  075020.
  [arXiv:0804.0717 [hep-ph]].
   
\bibitem{Allanach:2001kg}
  B.~C.~Allanach,
  Comput.\ Phys.\ Commun.\  {\bf 143} (2002) 305
  [arXiv:hep-ph/0104145].

\bibitem{arXiv:0902.4644}
  S.~Antusch and M.~Spinrath,
  Phys.\ Rev.\ D\ {\bf 79} (2009) 095004
  [arXiv:0902.4644 [hep-ph]].

\bibitem{Mahmoudi:2008tp}
  F.~Mahmoudi,
  Comput.\ Phys.\ Commun.\  {\bf 180} (2009) 1579
  [arXiv:0808.3144 [hep-ph]];
  F.~Mahmoudi,
  Comput.\ Phys.\ Commun.\  {\bf 178} (2008) 745
  [arXiv:0710.2067 [hep-ph]].
  
\bibitem{arXiv:1010.1589}
  D.~Asner {\it et al.} [Heavy Flavor Averaging Group Collaboration],
  arXiv:1010.1589 [hep-ex].

\bibitem{CMS-PAS-BPH-11-019}
  CMS and LHCb Collaborations,
  CMS-PAS-BPH-11-019, LHCb-CONF-2011-047.

\bibitem{PDG}
K.~Nakamura {\it et al.} [ Particle Data Group Collaboration ],
J.\ Phys.\ G {\bf G37}, 075021 (2010).
  
\bibitem{Profumo:2011zj}
  S.~Profumo,
  Phys.\ Rev.\  {\bf D84}, 015008 (2011).
  [arXiv:1105.5162 [hep-ph]].
  
\bibitem{Buchmueller:2011ki}
  O.~Buchmueller, R.~Cavanaugh, D.~Colling, A.~De Roeck, M.~J.~Dolan, J.~R.~Ellis, H.~Flacher, S.~Heinemeyer {\it et al.},
  Eur.\ Phys.\ J.\  {\bf C71 } (2011)  1722.
  [arXiv:1106.2529 [hep-ph]].  

\bibitem{prospino}
  W.~Beenakker, R.~Hopker, M.~Spira,
    [hep-ph/9611232].
    
\bibitem{arXiv:1111.3639}
  J.~S.~Gainer, R.~Huo and C.~E.~M.~Wagner,
  arXiv:1111.3639 [hep-ph].

\bibitem{HUTP-79-A026}
  H.~Georgi and C.~Jarlskog,
  Phys.\ Lett.\ B\ {\bf 86} (1979) 297.

      
\end{thebibliography}
\end{document}